\documentclass[aps,prl,reprint,twocolumn]{revtex4-2}
\usepackage{bbold}
\usepackage{amsmath,amsfonts,amsmath,mathtools,bbm,bm, amssymb}
\usepackage{graphicx}

\usepackage{ragged2e}
\usepackage{caption}
\DeclareCaptionJustification{justified}{\justifying}
\usepackage{subcaption, stackengine}

\usepackage{braket}
\usepackage[colorlinks,linkcolor=blue,urlcolor=blue,citecolor=blue]{hyperref}
\usepackage[all]{hypcap}

\usepackage{lipsum}
\usepackage{blindtext}
\usepackage{enumitem}
\usepackage{xcolor}
\usepackage[normalem]{ulem}
\usepackage{orcidlink}

\usepackage{float}
\setcitestyle{numbers,square}

\newcommand{\Tr}{\mathrm{Tr}}

\def\({\left (}
\def\){\right )}

\graphicspath{{Figure/PNG/}{Figure/PDF/}{Figure/EPS/}{Figure/TEX/}{Figure/}}

\newcommand\mydots{\hbox to 1em{.\hss.\hss.}}

\usepackage{titlesec}
\titleformat{\section}[runin]
  {\bf}{}{1em}{\phantomsection}[]
\titleformat{\subsection}[runin]
  {\it}{}{1em}{\phantomsection}[---]

\begin{document}

\title{Multipartite entanglement structure of monitored quantum circuits}

\author{Arnau Lira-Solanilla}
\affiliation{Institut f\"ur Theoretische Physik, Universit\"at zu K\"oln, Z\"ulpicher Stra{\ss}e 77, 50937 Cologne, Germany}

\author{Xhek Turkeshi}
\affiliation{Institut f\"ur Theoretische Physik, Universit\"at zu K\"oln, Z\"ulpicher Stra{\ss}e 77, 50937 Cologne, Germany}

\author{Silvia Pappalardi}
\affiliation{Institut f\"ur Theoretische Physik, Universit\"at zu K\"oln, Z\"ulpicher Stra{\ss}e 77, 50937 Cologne, Germany}

\date{\today}

\begin{abstract}
Monitored quantum circuits have attracted significant interest as an example of synthetic quantum matter, intrinsically defined by their quantum information content. 
Here, we propose a multipartite entanglement perspective on monitored phases through the lens of quantum Fisher information. 
Our findings reveal that unstructured monitored random circuits fail to exhibit divergent multipartite entanglement even at criticality, highlighting their departure from standard quantum critical behavior. 
However, we demonstrate that genuinely multipartite entangled phases can be realized through two-site measurements, provided a protection mechanism is in place. 
This work positions multipartite entanglement as a valuable perspective for the study of interacting monitored circuits and broader frameworks of noisy quantum dynamics. 
\end{abstract}

\maketitle

\textbf{Introduction}
Quantum computers, while not yet practical for solving real-world problems, are already redefining our understanding of many-body physics~\cite{preskill2018quantumcomputingin}. 
Indeed, these platforms host unique dynamical phenomena, often referred to as synthetic or monitored quantum matter, which are fundamentally described by quantum circuits composed of unitary gates and on-the-fly measurements~\cite{fisher2023random}. 
Unlike traditional phases of matter characterized by order parameters such as magnetization, these phases are instead defined by the quantum informational resources they process, offering a powerful framework to classify and understand many-body systems~\cite{gullans2020dynamical,gullans2021quantum,ippoliti2024learnability,agrawal2024observing,turkeshi2024error,lovas2024quantum,ferte2024solvable}.

A pivotal role in our understanding of these phases is played by bipartite entanglement, often quantified by the entanglement entropy, which measures the number of distillable Bell pairs shared between two parties~\cite{amico2008entanglement,horodecki2009quantum,calabrese2009entanglement,laflorencie2016quantum}. 
Bipartite entanglement has been instrumental in revealing the phenomenon of measurement-induced transitions, where the scaling of entanglement undergoes a qualitative change~\cite{noel2022measurement,koh2023measurement,hoke2023measurement}.
For unstructured monitored circuits~\cite{potter2022entanglement,lunt2022quantum}, when local measurements are dynamically applied within the circuit at a given rate, the bipartite entanglement of individual trajectories transitions from an extensive regime (volume-law), where entanglement grows proportionally with system size, to a sub-extensive regime (area-law)~\cite{li2018quantum,li2019measurement,skinner2019measurement,jian2020measurement,bao2020theory}.
However, bipartite entanglement alone cannot fully capture the complexity of these dynamical phases. 
A broader characterization using complementary quantum resources is essential for both advancing the classification of quantum many-body systems and linking them to potential quantum applications~\cite{chitambar2019quantum}.
Previous studies have demonstrated that coherence~\cite{sierant2022universal,shane2023coherence,colmenarez2024accurate,eckstein2024robust,venn2023coherent} and magic resources~\cite{niroula2024phase,bejan2024dynamical,fux2024entanglement,paviglianiti2024estimatingnonstabilizernessdynamicssimulating,gu2024magicinducedcomputationalseparationentanglement,cheng2024emergentunitarydesignsencoded} provide substantial insights on these systems, even pointing to applications on their observability~\cite{gullans2020scalable,li2023cross,kamakari2024experimentaldemonstrationscalablecrossentropy}. 

\begin{figure}[t!]
    \centering
    \includegraphics[width=0.45\textwidth]{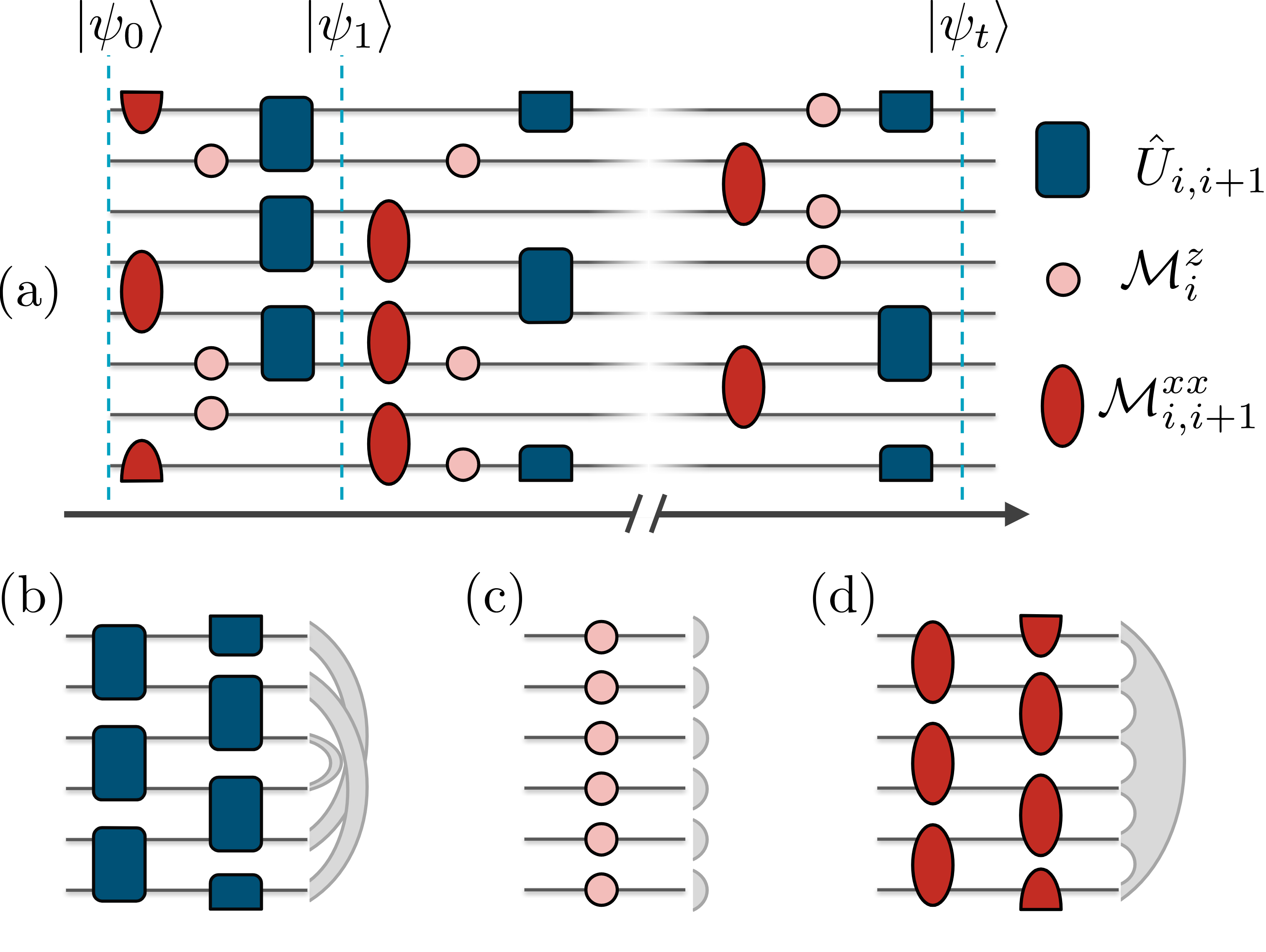}
    \caption{(a) Schematic description of a particular disorder realization of the circuit structure. The system is evolved for long times $t\sim O(L)$, in the horizontal axis, to reach a state with stationary entanglement properties. (b)-(d) In gray, typical entanglement structure generated by the different operations present in the circuit: random two-qubit gates $\hat U_{i, i+1}$, $\hat \sigma_i^z$ or $\hat \sigma_i^x \hat \sigma^x_{i+1}$-measurements respectively.
    }
    \label{fig:circuit}
\end{figure}

In this work, we take a fundamentally different approach by examining the \emph{multipartite entanglement structure} of different monitored quantum circuits, depicted in Fig.\ref{fig:circuit}. Multipartite entanglement captures the shared quantum correlations across multiple subsystems and is often quantified using the quantum Fisher information ~\cite{braunstein1997statistical, paris2009quantum, pezze2009entanglement, giovannetti2011advances, hylluys2012fisher, toth2012multipartite, 
toth2014quantum,  pezze2014quantumtheoryphaseestimation,pezze2018quantum, liu2019quantum, yu2022quantum}.
This quantifier not only offers a deeper understanding of the nature of quantum correlations for pure and mixed states but also directly links to practical applications in quantum metrology, where it serves as a key indicator of quantum advantage in precision phase estimation tasks. 

Our findings reveal a surprising property of unstructured monitored circuits: \emph{multipartite entanglement is entirely absent at any measurement rate}. In other words, monitored systems with local measurements are confined to a trivial multipartite phase. 
This behavior starkly contrasts with the phenomenology observed in fine-tuned models such as non-interacting fermions~\cite{paviglianiti2023multipartite} and semiclassical monitored systems~\cite{passarelli2024manybody,poggi2024measurementinduced}, where a sharp transition occurs between an extensive and trivially multipartite entangled phase. 
On the other hand, structured circuits as those with a global conservation law can host genuinely multipartite entangled phases, highlighting the critical role of a protection mechanism for generating and stabilizing multipartite entanglement.

\textbf{Multipartite entanglement and quantum Fisher information.} 
Multipartite entanglement extends beyond the simpler notion of bipartite entanglement, which quantifies the number of Bell pairs shared between two parties, to describe the complex quantum correlations distributed among multiple parties in a system. 
Several frameworks exist to quantify multipartite entanglement.
Throughout this work, we use the quantum Fisher information (QFI) $\mathcal{F_Q}$, which gives direct access to the size of the entangled blocks for pure and mixed states and plays a fundamental role in quantum metrology~\cite{paris2009quantum, pezze2009entanglement, pezze2018quantum}.
For a pure state $\ket{\psi}$ and operator $\hat O$, the QFI is given by ${\mathcal{F_Q}(\hat{O}) = 4 \left ( \langle \hat{O}^2\rangle - \langle \hat{O} \rangle^2 \right )}$, with $\langle \hat O\rangle\equiv \langle \psi |\hat O |\psi\rangle$~\cite{braunstein1997statistical}.

Particularly relevant is when the observable $\hat O$ is a collective operator ${\hat O = \frac 12\sum_{i=1}^L \hat o_i}$, that is a sum of locally supported operators $\hat{o}_i$. In this case, the QFI exhibits key mathematical properties, such as additivity and convexity~\cite{toth2014quantum,pezze2018quantum}, making it a valuable tool for probing the multipartite entanglement structure of $L$-qubits states~\cite{giovannetti2011advances, hylluys2012fisher, toth2012multipartite}. 
If the QFI density satisfies~\cite{toth2012multipartite, hylluys2012fisher}
\begin{equation}
    f_\mathcal{Q}(\hat{O}) \equiv \frac{\mathcal{F_Q}(\hat{O})}{L} > m,
\end{equation}
for $m<L$ integer, then at least $m+1$ parties in the system are mutually entangled. 
In particular, when ${f_Q\propto L}$, all qubits are clustered in one entangled block, and the state is referred to as \emph{genuinely multipartite}. 
The most notable example for these type of states is given by the Greenberger-Horne-Zeilinger state $|\mathrm{GHZ}\rangle=(|0\rangle^{\otimes L}+|1\rangle^{\otimes L})/\sqrt{2}$~\cite{greenberger2007goingbellstheorem}, which presents maximal QFI $\mathcal{F_Q}(\hat O=\frac 12 \sum_i \hat\sigma_i^z) = L^2$~\footnote{Throughout this work, the Pauli matrices acting on site $i$ are denoted as $\hat \sigma_i^\alpha$ with $\alpha=x,y,z$.}. This state is known to have finite connected correlations at arbitrary distances, a condition known as long-range order~\cite{yang1962concept}. 
By the invariance of QFI under local unitary transformation, any state 
\begin{equation}
    \otimes_{j=1}^L \hat{U}^{(n_j)}_j |\mathrm{GHZ}\rangle = \frac{|n_1 n_2\dots n_L\rangle+ |\bar{n}_1 \bar{n}_2\dots \bar{n}_L\rangle}{\sqrt{2}}
    \label{eq:invariance_qfi}
\end{equation}
remains genuinely multipartite entangled and also referred to as GHZ or cat states. 
Here, the $\hat{U}^{(n_j)}_j$ are onsite unitary operators that transform $\hat{U}^{(n_j)}_j|0\rangle = |n_j\rangle$ and $\hat{U}^{(n_j)}_j|1\rangle = |\bar{n}_j\rangle$, and the operator of interest becomes $\hat{O}_{\{\mathbf{n}\}} =\sum_{i=1}^L \mathbf{n}_i\cdot \hat{\boldsymbol{\sigma}}_i/2$, 
where the local directions $\mathbf{n}_i$ are fixed by the condition $\mathbf{n}_i\cdot \hat{\boldsymbol{\sigma}}_i=\hat{U}^{(n_i)}_i \hat{\sigma}^z_i (\hat{U}_i^{(n_i)})^\dagger$. 

From the above discussion, it is clear that different operators $\hat{O}$ yield different bounds, and there is no systematic method -- beyond explicit knowledge of the system’s physics~\cite{hauke2016measuring, pezze2017multipartite} -- to identify the optimal one. 
Therefore, in the absence of prior information, the multipartite entanglement structure is estimated by maximizing the QFI of collective operators $\hat{O}_{\{\mathbf{n}\}}$ over the directions $\{\textbf{n}_i\}$ 
\begin{equation}
    \label{eq_qfi_max}
   \mathcal{F_Q} \equiv \max_{\mathbf{n}_j} \mathcal{F_Q}(\hat{O}_{\{\mathbf{n}\}})=\max_{\mathbf{n}_j}\left\{\sum_{\alpha,\beta}\sum_{i,j}n_i^\alpha n_j^\beta C_{ij}^{\alpha\beta}\right\} ,
\end{equation}
where $C_{ij}^{\alpha\beta} = \langle \hat{\sigma}_i^\alpha \hat{\sigma}_j^\beta\rangle- \langle\hat{\sigma}_i^\alpha\rangle \langle \hat{\sigma}_j^\beta\rangle$ is the connected correlation function between two sites $i, j$.

Importantly, the quantum Fisher information has attracted significant attention in the many-body context due to its connection to susceptibilities~\cite{hauke2016measuring, pappalardi2017multipartite, brenes2020multipartite}. Not only does this make the QFI experimentally accessible~\cite{laurell2021quantifying, scheie2021witnessing, vitale2023robust}, but also leads to a characteristic universal divergence at quantum critical points~\cite{hauke2016measuring, frerot2018quantum, rams2018at, yaoming2021dynamic}.

\begin{figure}[t!]
    \centering
    \begin{subfigure}[t]{0.49\columnwidth}
        \phantomcaption
        \stackinset{l}{5pt}{t}{-5pt}{(\thesubfigure)}{
        \includegraphics[width=\textwidth]{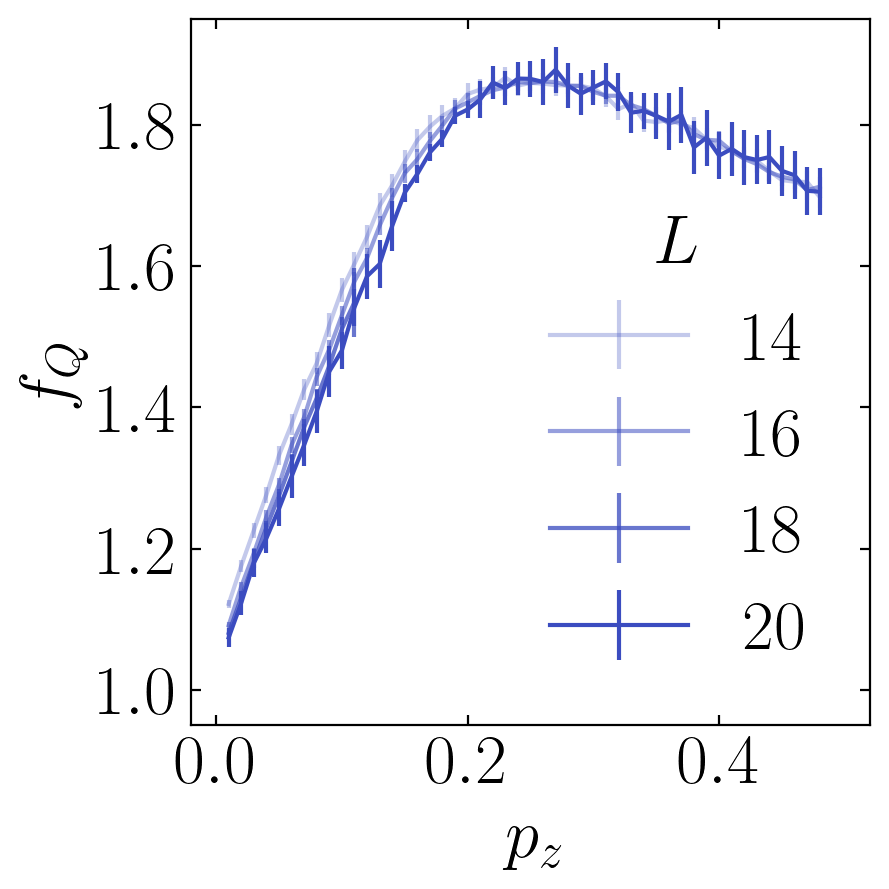}
        }   
        \label{fig:haar_rqc}
    \end{subfigure}
    \begin{subfigure}[t]{0.49\columnwidth}
        \phantomcaption
        \stackinset{l}{5pt}{t}{-5pt}{(\thesubfigure)}{
        \includegraphics[width=\textwidth]{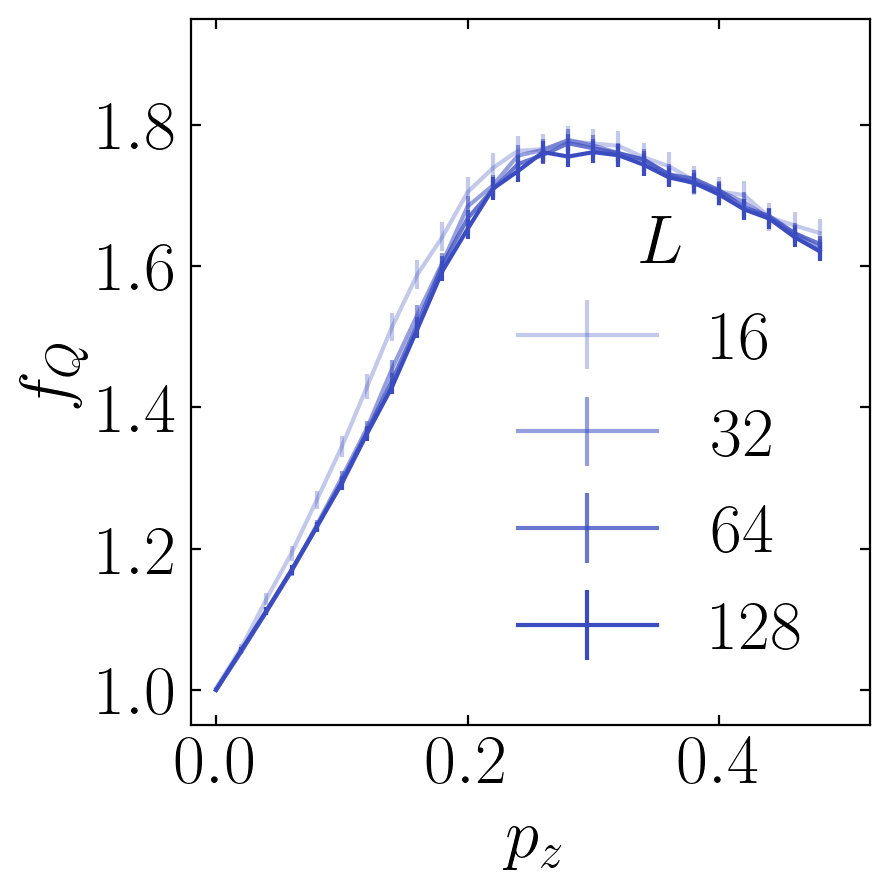}
        }  
        \label{fig:clifford_rqc}
    \end{subfigure}
    \vspace{-5mm}
    \caption{Average stationary state QFI for Haar (a) and stabilizer (b) random quantum circuits in the presence of onsite measurements for different system sizes $L$ and measurement rates $p_z$. 
    In both setups, larger system sizes converge to a constant value $f_\mathcal{Q}<2$, signaling the absence of multipartite entanglement beyond two parties. 
    }
    \label{fig:ensembles_rqc}
\end{figure}

\textbf{Unstructured monitored circuits.}
We start our analysis investigating the multipartite entanglement structure of the standard models of measurement-induced phase transitions: unstructured monitored random circuit~\cite{fisher2023random,li2018quantum,skinner2019measurement}.
Each time step consists of a layer of measurements of the operator $\hat{\sigma}^z_j$, performed on each site $j=1,2,\dots,L$ with probability $p_z$ and a layer of nearest neighboring gates $\hat{U}_{j,j+1}$ in a brickwork pattern, cf.~Fig.\ref{fig:circuit}, uniformly sampled from the full unitary group or the Clifford subgroup for Haar and Clifford circuits, respectively. 
As a result, the evolution is stochastic and fixed by quantum trajectories $|\psi_\mathbf{m}\rangle$ specified by the realization $\mathbf{m}$ of the circuit. 
Extensive numerical studies have established the presence of a measurement-induced phase transition between a volume-law bipartite entangled phase and an area-law phase in the stationary regime of typical trajectories~\cite{zabalo2020critical, zabalo2022operator, sierant2022measurement, sierant2022universal}. The critical points are approximately  $p^\mathrm{H}_z \simeq 0.17$  for Haar circuits and  $p^\mathrm{C}_z \simeq 0.16 $ for Clifford circuits, which are governed by two distinct conformal field theories~\cite{jian2020measurement, li2021conformal}.

Hereby, we study the quantum Fisher information of typical trajectories in both Haar and Clifford circuits. 
Starting from the state $|\psi_0\rangle=|0\rangle^{\otimes L}$, we compute the time evolved state $|\psi_{\mathbf{m}_t}\rangle$ at circuit depth $t=4L$, where the system reaches a stationary regime~\cite{zabalo2020critical}.
For each choice of parameters $L$ and $p_z$ we take at least $10^3$ realizations of quantum trajectories.
For each realization, we compute the QFI density $f_\mathcal{Q}(\mathbf{m}_t)$ by optimizing over the set of local directions  $\{\mathbf{n}_k\}$ using a \emph{classical annealing algorithm}. 
As detailed in the End Matter, the problem is a ground state search of the classical Hamiltonian $\mathcal{H}=-f_\mathcal{Q}(\hat{O}_{\{\mathbf{n}_k\}})$, where $\{\mathbf{n}_k\}$ in Eq.~\eqref{eq_qfi_max} are interpreted as spin variables and the trajectory correlation functions $C_{ij}^{\alpha\beta}$ as interaction terms \cite{paviglianiti2023multipartite}. 
The typical QFI is obtained averaging over the realizations, which, with a slight abuse of notation, we denote $f_\mathcal{Q}\equiv \mathbb{E}_\mathbf{m}[f_\mathcal{Q}(\mathbf{m}_t)]$.
The results are shown in Fig.\ref{fig:haar_rqc} and Fig.\ref{fig:clifford_rqc} for Haar and Clifford circuits, respectively. 

Remarkably, despite the divergent correlation length and its associated logarithmic behavior for the bipartite entanglement entropy~\cite{zabalo2020critical,sierant2022measurement}, we find that the asymptotic state displays no divergent behavior of multipartite entanglement at the critical point. 
Most strikingly, by increasing system sizes $L$, the QFI density does not present any feature upon increasing the system size, manifesting at most entangled blocks of size two, i.e.
\begin{equation}
    \lim_{L\to \infty}  f_\mathcal{Q}<2 \;\;\; \text{for any } p_z\in [0,1].\label{eq:ma}
\end{equation}

Off-criticality, we expect this result because the QFI is a sum of connected correlation functions, which decay exponentially with distance when a finite correlation length is present \cite{desaules2022extensive}. As a result, the QFI density remains finite.
What is more surprising is that the Fisher density is finite even at the critical point of bipartite entanglement, in stark contrast with what occurs in critical non-interacting monitored phases~\cite{paviglianiti2023multipartite}. 
We can present a heuristic justification following the analysis for the critical phases of clean systems~\cite{hauke2016measuring}.
General arguments of scaling invariance predict an algebraic decay of connected correlation functions, $C_{i,j}^{\alpha\beta}\sim |i-j|^{-2\Delta_{\alpha,\beta}}$ leading to 
\begin{equation}
    f_\mathcal{Q}\sim a+ b L^{1-2\Delta}\ ,
    \label{eq:qfi_scaling}
\end{equation}
where $a,\ b$ are system size independent constants and $\Delta$ is the operator scaling dimension $\Delta_{\alpha,\beta}$ for which the algebraic decay is slowest. 
Within this picture, the Fisher density can diverge with system size only for $\Delta <1/2$. 
For the circuit under consideration, the precise identification of the connected correlations depends crucially on the order of the optimization and trajectory average~\footnote{
Crucially, the computation remains non-trivial even in the limit  $p \to 0 $, corresponding to purely unitary dynamics  $|\psi_t\rangle = \mathcal{U}|\psi_0\rangle$  for some unitary operator  $\mathcal{U} $. Specifically, the local rotations depend on the evolution operator  $U^{(n_j)}_j(\mathcal{U}) $, and averaging over circuit realizations is non-linear, cf. also Ref.~\cite{paviglianiti2023multipartite}.
A similar reasoning holds also in the limit $p\to 1$.
}. 
Neglecting this point, we can estimate the behavior of the QFI density at criticality inserting the estimated $\Delta\sim 2$ obtained via mutual information in Ref.~\cite{skinner2019measurement, li2019measurement}. This fact signals a decay in space which is too fast to induce growing multipartite entanglement and leads to $f_\mathcal{Q}\sim O(1)$, consistent with our data. 

To summarize, locally monitored random circuits are characterized by purely bipartite entanglement and intensive multipartiteness. 
In other words, standard monitored phases have trivial multipartite entanglement and are metrologically useless \cite{paris2009quantum, pezze2009entanglement, pezze2018quantum}. 
We conclude by noting that the heuristic picture may also explain the trivially multipartite phase at low measurement rate in 
Ref.~\cite{paviglianiti2023multipartite}. Since the effective conformal field theory is non-unitary \cite{chen2020emergent}, the operator scaling can flow to $\Delta<1/2$ under a threshold measurement rate and the system loses its multipartite entanglement structure.

\begin{figure}[t]
    \centering
    \includegraphics[width=0.45\textwidth]{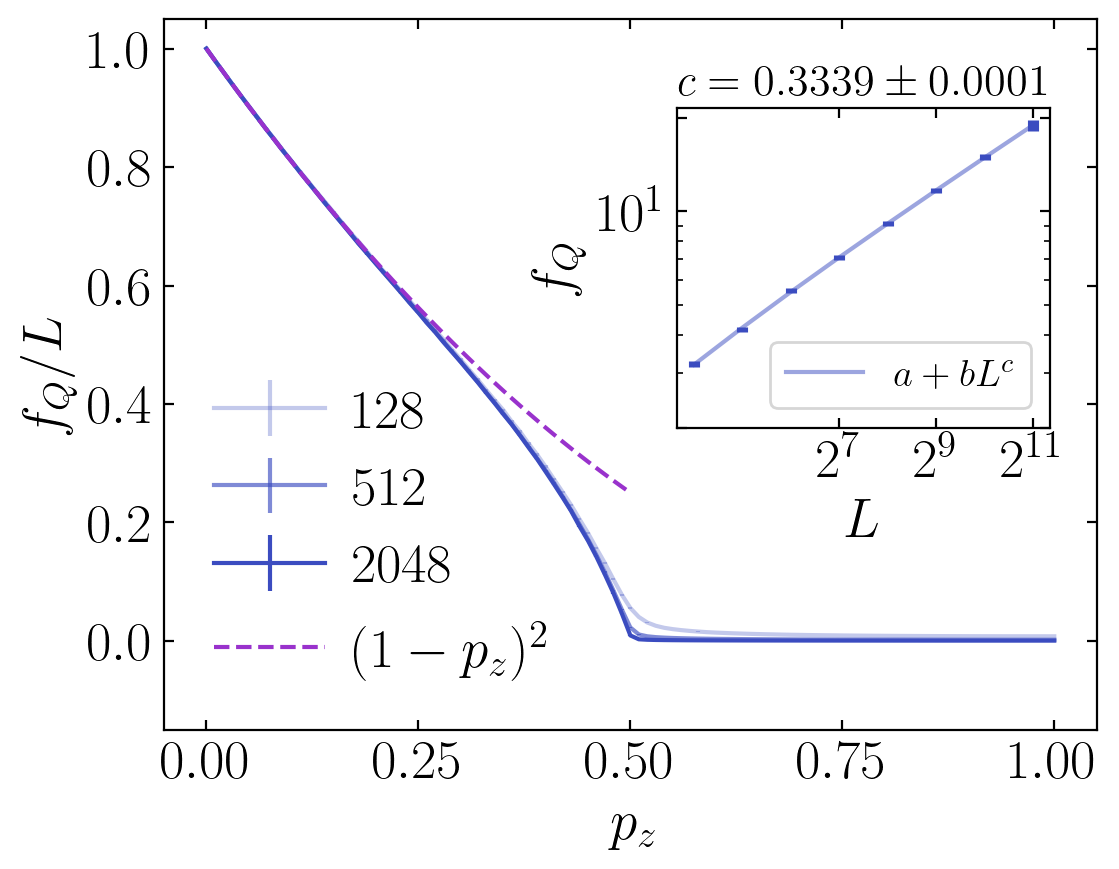}
    \caption{QFI as a function of the single-qubit measurement $p_z$ in the projective Ising model. Inset: scaling of the QFI at criticality, $p_z=0.5$. }
    \label{fig:pu000}
\end{figure}

\textbf{Genuinely multipartite monitored circuits.}
In contrast to onsite measurements, which can only decrease multipartite entanglement by factorizing the system, measurement on Pauli operators acting on two or more sites can lead to \emph{macroscopic cat states}, i.e., genuinely multipartite states.  
For concreteness, we consider an initial state $|\psi_0\rangle$ and randomly act with measurements on neighboring $\hat{\sigma}^x_i \hat{\sigma}^x_{i+1}$ with $i=1,\dots,L-1$. 
These dynamics describe a simple stochastic process, fixed by the measurement projectors $\Pi^{\pm}_{i,i+1}=(\hat{\mathbb{I}}\pm\hat{\sigma}^x_i \hat{\sigma}^x_{i+1})/2$. 
Irrespective of the measurement results, these operators either generate an entangled pair when acting on $\ket{00}$, or enlarge a given entangled cluster absorbing a qubit in the state $\ket{0}$, or coalesce separate clusters by entangling two of their qubits~\cite{lang2020entanglement}. A complete account of these rules is detailed in End Matter. 
The stationary state of these dynamics results in a state of the form Eq.~\eqref{eq:invariance_qfi}, with the individual $|\theta_i\rangle=|\pm_i\rangle$ randomly sampled, cf. Fig.\ref{fig:circuit}(d).

How stable is such a multipartite entangled state, created in this manner, under competing measurements or unitary gates? 
To gain insights into the problem, we first study the QFI's behavior under a circuit composed of measurements on $\hat{\sigma}^x_i \hat{\sigma}^x_{i+1}$ and on $\hat{\sigma}^z_i$, the so-called projective Ising model~\cite{lang2020entanglement, tarabunga2024magictransitionmeasurementonlycircuits}. 
This model is mappable to a geometric problem, allowing us to obtain an intuitive interpretation of the QFI. In a second step, we will incorporate the effect of unitary gates.

The dynamics of the projective Ising model are generated by the competition of interspersed layers of $\hat{\sigma}^x_i \hat{\sigma}^x_{i+1}$ measurements, occurring on each site with probability $p_{xx}=1-p_z$, and local $\hat{\sigma}^z_i$ drawn with probability $p_z$, cf. Fig.\ref{fig:circuit}.
In essence, the dynamics follow the update rule of the projector $\Pi_{i,i+1}^\pm$ together with a new rule accounting for the projectors onto the computational basis $|0\rangle,\ |1\rangle\}$. 
This set of simple rules allows to map the model to the bond percolation problem on a square lattice, with percolation probability fixed by $p_z$ and presenting a critical point at $p_z^c=1/2$~\cite{essam1980percolation}.

Crucially, this interpretation is tied to a fast simulation of the dynamics with $O(N)$ resources~\cite{lang2020entanglement}. 
This fact also applies to the dynamics of multipartite entanglement: at each time-step, the system is in a tensor product of cat states~\eqref{eq_qfi_max} of different sizes and of product states, cf. End Matter for details.  
As a result, no optimization is required to compute the quantum Fisher information of this model, which takes a compelling geometrical interpretation.
By additivity, the QFI is fixed by the number $l_s$ of unentangled qubits and the contribution of GHZ clusters, denoted by $c\in \mathcal{C}$ and contributing to $l_c^2$ multipartite-ness, resulting in
\begin{equation}
    \label{eq_qfi_loop}
    \mathcal{F}_\mathcal{Q} = l_s + \sum_{c\in\mathcal{C}} l_c^2\ .
\end{equation}

The efficient simulability of the model, combined with the expression in Eq.\eqref{eq_qfi_loop}, enables us to compute the optimal QFI for systems up to  $L = 2048$.
For  $p_z < p_z^c$, the stationary multipartite entanglement grows extensively with the system size  $f_\mathcal{Q} \propto L$, indicating a stable and genuinely multipartite phase. 
This phase can be understood in $p_z\simeq 0$ from the percolation mapping. In this regime of rare $\hat{\sigma}_i^z$ measurements, we can approximate the system as in a unique cat state up to perturbative corrections in $p_z$. This leads to 
\begin{equation}
    \label{eq_fQ_approx}
    \lim_{L\to \infty} \frac{f_\mathcal{Q}}{L} = (1-p_z)^2\;,
\end{equation}
that, as shown in Fig.\ref{fig:pu000}, capture the qualitative physics as small $p_z$. 
Conversely, for  $p_z > p_z^c$, the state becomes separable. This behavior is illustrated in Fig.\ref{fig:pu000}, which shows the rescaled QFI,  $f_\mathcal{Q}/ L$, as a function of  $p_z$ for different system sizes. At criticality,  $p_z = p_z^c$, the QFI exhibits universal divergence $f_\mathcal Q\sim L^{1/3}$, as shown in the inset of Fig.\ref{fig:pu000}.
Since the scaling dimension at percolation is $\Delta = 1/3$~\cite{lang2020entanglement}, this result also matches the prediction of Eq.~\eqref{eq:qfi_scaling}.

The phase transition corresponds to the spontaneous symmetry breaking of the $\mathbb{Z}_2$ parity symmetry, ${\hat{S} = \prod_{i=1}^L \hat{\sigma}_i^z}$, which commutes with both measured Pauli strings. As we will now discuss, this symmetry is central to the stability of the multipartite phase. 

To illustrate this fact, we now discuss the resilience of the multipartite entanglement phase upon introduction of unitary gates. 
Concretely, we modify the circuit as in Fig.\ref{fig:circuit} where on top of $\hat{\sigma}^x_i \hat{\sigma}^x_{i+1}$ ($\hat{\sigma}^z_i$) measurements with probability $p_{xx}$ ($p_z$), we also apply nearest neighboring random unitaries $\hat{U}_{i,i+1}$ with probability $p_u$ bounded by $p_u + p_z + p_{xx}=1$. 
If $[\hat{U}_{i,i+1},\hat{S}]\neq 0$ the system multipartite phase is completely lost for any $p_u>0$, cf. End Matter.
On the other hand, when the unitary gates preserve the symmetry, the genuinely multipartite phase extends to finite $p_u>0$ up to a critical threshold $p_u^c(p_z)$. 
In essence, \emph{symmetry acts as a fundamental protection mechanism to stabilize genuinely multipartite entanglement phases in monitored systems}.

\begin{figure}[t]
    \centering
    \includegraphics[width=0.46\textwidth]{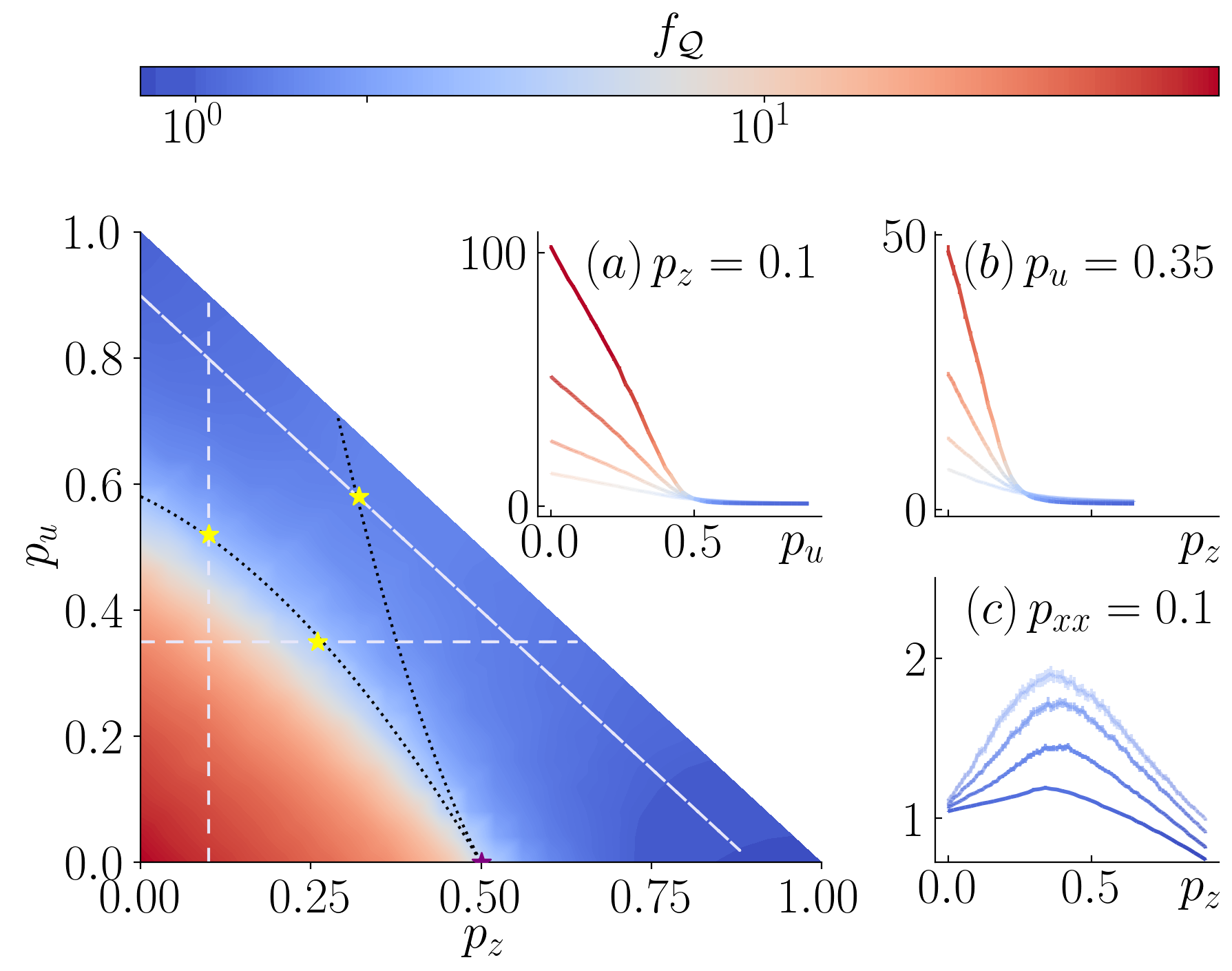}

    \caption{(main) Phase diagram characterizing the multipartite entanglement structure of circuits parametrized by $(p_z,p_u)$ and $L=64$. (a)-(c) Quantum Fisher information density as a function of the single-qubit measurement $p_z$ and random unitary $p_u$ rates, given a fixed $p_z$, $p_u$, and $p_{xx}$ respectively, for system sizes $L=16, 32, 64, 128$ indicated by an increase of opacity. They are the cross sections of the straight dashed lines in the main figure, the position of the critical points is marked with a star.}
    \label{fig:phase_diagram}
\end{figure}

Upon inclusion of random unitary gates, the geometric picture of the multipartite information~\eqref{eq_qfi_loop} is lost and the phase diagram for generic $p_u$ and $p_z$ requires optimizing over directions, cf. Eq.~\eqref{eq_qfi_max}.
Our numerical results for system sizes up to $L\le 128$ averaging over at least $3000$ realizations are summarized in Fig.\ref{fig:phase_diagram}. 
From the multipartite entanglement perspective, we identify two clearly distinguished phases: a genuinely multipartite phase $f_\mathcal{Q} \propto L$ below the critical line $p_u^c(p_z)$, and a trivial one for $f_\mathcal{Q} = O(1)$.

It is indicative to contrast these results with the bipartite entanglement phases, extracted in the End Matter via the tripartite mutual information~\cite{zabalo2020critical}. 
Bipartite entanglement hosts a volume-law and two area-law phases characterized by short-range and long-range order, represented in Fig.\ref{fig:phase_diagram} by dashed separation lines. 
We focus our attention on the behavior of the quantum Fisher information for $L=16, 32, 64, 128$ on three lines crossing these bipartite critical lines (yellow stars in Fig.\ref{fig:phase_diagram}). 

In Fig.\ref{fig:phase_diagram}(a), we show the $f_\mathcal{Q}$ versus $p_u$ at constant $p_{z}=0.1$.
This setup presents a transition between a genuinely multipartite entangled phase $f_\mathcal{Q}\propto L$ to a constant value $f_\mathcal{Q} = O(1)$. Crucially, the critical point coincides
with the bipartite entanglement transition at $p_u^c\sim 0.52$.\\
In Fig.\ref{fig:phase_diagram}(b), we show the $f_{\mathcal{Q}}$ as a function of $p_z$ at constant $p_{u}=0.35$, which also hosts two phases of multipartite entanglement with extensive and intensive $f_{\mathcal{Q}}$, below and above $p_z^c\sim 0.26$. 
Interestingly, while the bipartite entanglement undergoes a second transition between a volume-law and a short-range area-law phase at $p_z^{c'}\sim 0.38$ (cf. End Matter), this 
occurs without any noticeable change from the perspective of multipartite entanglement. This aligns with our previous findings [cf. Eq.\eqref{eq:ma}], which indicates no evidence of extensive growth in $f_\mathcal{Q}$ across such transitions. \\
In Fig.\ref{fig:phase_diagram}(c), we show the $f_\mathcal{Q}$ as a function of $p_z$ at constant $p_{xx}=0.1$. Again, the volume-to-area law bipartite entanglement transition corresponds to a unique and trivial multipartite phase, always bounded for any $L\le 128$ by a constant value, similarly to what occurs for random unstructured systems, cf. Eq.~\eqref{eq:ma}. 
Lastly, analyzing the behavior at the critical points of Fig.\ref{fig:phase_diagram}(a,b) we find that the QFI density \emph{does not scale with system size}, namely $f_{\mathcal{Q}}(p^c)= O(1)$ for any $p_u^c(p_z)>0$. This result is consistent with the $\Delta\simeq 0.7$ extracted for similar models in Ref.~\cite{sang2021entanglement} once included in the scaling hypothesis Eq.~\eqref{eq:qfi_scaling}.

\textbf{Conclusions.}
We studied the multipartite entanglement of monitored quantum circuits.
Unstructured random circuits display only a trivial multipartite entangled phase, in stark contrast with the bipartite entanglement that develops volume-to-area transitions. 
Stabilizing a genuinely multipartite monitored phase requires a protection mechanism: we provided a concrete example focusing on parity-symmetric random circuits. 
Despite the divergence of the correlation length at criticality,  we have shown that the scaling exponent of connected correlation functions is the relevant figure that controls how the quantum Fisher information scales with system size. 

We envision several follow-ups. 
First, our work reveals an intuitive geometric interpretation of quantum Fisher information in the projective Ising model~\cite{lang2020entanglement}. 
This opens exciting opportunities for the theoretical exploration upon inclusion of structured unitary gates and measurements, more prominently in Majorana loop models~\cite{nahum2020entanglement,klocke2022topological,klocke2023majorana,merritt2023entanglement,loio2023purification,klocke2024powerlaw,klocke2024entanglementdynamicsmonitoredkitaev} that present a clear geometrical interpretation~\cite{nahum2013length,nahum2013loop,nahum2015deconfined}.
Additionally, the observation that QFI in monitored systems acts as an indicator of purely long-range order calls for a general understanding of its role in characterizing topological phases at equilibrium~\cite{pezze2017multipartite, zhang2018characterization, zhang2022multipartite} and beyond~\cite{lavasani2021measurement,lavasani2021topological,sang2021measurementprotected,kells2023topological,lavasani2023monitored,sriram2023topology,zhu2023qubitfractionalizationemergentmajorana}. 
More broadly, the quantum Fisher information enables the study of the entanglement structure in mixed states, opening new avenues of research and the possibility of tackling open problems in monitored quantum dynamics beyond pure systems.

\begin{acknowledgments} 
\textit{Acknowledgments.} We thank A. Paviglianiti for helpful exchanges regarding the simulated annealing code, and M. Buchhold, L. Piroli and P. Sierant for discussions. We acknowledge funding by the Deutsche Forschungsgemeinschaft (DFG, German Research Foundation) under Projektnummer 277101999 - TRR 183 (project B01 and B02), and under Germany's Excellence Strategy - Cluster of Excellence Matter and Light for Quantum Computing (ML4Q) EXC 2004/1 - 390534769. The most demanding numerical simulations were performed on the CHEOPS cluster at RRZK Cologne.

\textit{Data Availability.} The code of our data is available at Ref.~\cite{dataavail}. 
\end{acknowledgments}

\bibliography{biblio}
\bibliographystyle{apsrev4-2}

\clearpage

\setcounter{section}{0}
\setcounter{secnumdepth}{2}

\begin{center}
    \textbf{End Matter}
\end{center}

\textbf{Maximum QFI via annealing}

We show the steps to find the observable $\hat{O}$ that maximizes the QFI. Eq.~\ref{eq_qfi_max} is interpreted as the ground-state search of the classical Hamiltonian $H_\text{cl}(\{\mathbf{n}\})=-\sum_{\alpha,\beta}\sum_{ij}n_i^\alpha n_j^\beta C_{ij}^{\alpha\beta}$. The unit vectors, understood as classical spin states in a particular configuration, are parametrized as
\begin{equation}
    \mathbf{n}_k
    =\bigg( \sin(\theta_k)\cos(\phi_k), \; \sin(\theta_k)\sin(\phi_k), \; \cos(\theta_k)\bigg),    
\end{equation}
and optimized over the variables $\{\theta_k,\phi_k\}_{k=0}^{L-1}$. 

At each disorder realization of the circuit, we calculate all correlations $C_{ij}^{\alpha\beta}$ of the final state. Given a cooling schedule with temperatures $\{T_0, T_1, \dots, T_f\}$ and counts of iterations $\{i_0, i_1, \dots, i_f\}$, the simulated annealing algorithm is based on performing a Metropolis algorithm at each temperature, storing the configuration, and lowering the temperature until the ground state is found, cf. Ref.~\cite{landau2015computational}. The pseudo-code of the algorithm is the following:

1. Starting at $T_0$:
    
     (a) Generate a random initial configuration $\{\mathbf{n}_k^{i}({\theta}_k, {\phi}_k)\}_{k=1,\dots,L}$.
     
     (b) Calculate the cost function value $H_\text{cl}(\{\mathbf{n}^i\})$.
     
     (c) Choose a neighbor of the previous configuration, $\mathbf{n}^n= \mathbf{n}^i + \delta\mathbf{n}$, where $\delta\mathbf{n}=\delta \phi (\partial \mathbf{n}/\partial\phi) + \delta \theta (\partial \mathbf{n}/\partial\theta)$ and  $|\delta\mathbf{n}|\ll 1$.
        
     (d) Calculate the new cost function value $H_\text{cl}(\mathbf{n}^n)$.
     If $H_\text{cl}(\mathbf{n}^n) < H_\text{cl}(\mathbf{n}^i)$: accept the change and keep configuration.
        Otherwise, accept the change with probability $p=e^{-\Delta H_\text{cl}/T_0}$.
     
     (e) Repeat step 1c $i_0$ times to find $\mathbf{n}^\text{opt}_{T_0}$

2.  Loop through the pairs $\{(T_1, i_1)\dots (T_f, i_f)\}$ performing steps 1b-1f, starting from the optimized configuration.

3. After optimizing at $T_f$ and finding $\mathbf{n}^\text{opt}_{T_f}$, return the cost function  $H_\text{cl}(\mathbf{n}^\text{opt}_{T_f})$.

4. The QFI of the given disorder realization simulation is $\mathcal{F_Q} = - H_\text{cl}(\mathbf{n}^\text{opt}_{T_f})$.

We refer to the simulated annealing code for exact details on the chosen cooling schedule, where we loop through sets of $O(10^3)$ iterations with temperatures from $1$ to $0.02$, reducing the change $\delta$ in randomly sampled $\theta_k,\phi_k$ from $3\pi/2$ to $\pi/4$.

\textbf{Entangled states in the projective Ising model.}
Independently of the measurement outcome, a $\hat{\sigma}^x_i\hat{\sigma}^x_{i+1}$-measurement on initially separable qubits in the computational basis creates a Bell pair in the $\{|\pm\rangle\}$ basis. 
Without loss of generality, consider $\ket{00}$ the state of the $(1,2)$ qubits. 
We have
\begin{equation}
    \Pi_{1,2}^\pm\ket{00} = \Pi_{1,2}^\pm \big(\ket{++}+\ket{+-}+\ket{-+}+\ket{--}\big)/4
\end{equation}
gives rise after normalization to $(\ket{++}+\ket{--})/\sqrt{2}$ or $(\ket{+-}+\ket{-+})/\sqrt{2}$ with equal probability. Both Bell states present equivalent QFI as argued in Eq.\eqref{eq:invariance_qfi}. 

Consider instead a $\hat{\sigma}^x_i\hat{\sigma}^x_{i+1}$-measurement acting on a Bell pair and on an unentangled qubit, which for reference we consider respectively on qubits (1,2) and qubit 3. Applying the appropriate projector (given again by Born rule) $\Pi_{2,3}^\pm\left((\ket{++}+\ket{--}) \otimes \ket{0} \right)$ and renormalizing, we have $(\ket{+++}+\ket{---})/\sqrt{2}$ or $(\ket{++-}+\ket{--+})/\sqrt{2}$ with equal probability. 
Similarly, when a $\hat{\sigma}^x_i\hat{\sigma}^x_{i+1}$-measurement occurs between qubits in different Bell pairs, which for reference are located on qubits (1,2) and (3,4), 
$\Pi_{3,4}^\pm\left((\ket{++}+\ket{--}) \otimes (\ket{+-}+\ket{-+})\right)$ gives rise to $(\ket{+++-}+\ket{---+})/\sqrt{2}$ or $(\ket{++-+}+\ket{--+-})/\sqrt{2}$. 
These rules generalize iteratively to larger entangled clusters, substituting Bell pairs with GHZ states of larger qubit size. 
The operation creates multipartite entangled clusters, as illustrated in Fig.\ref{fig:circuit}d. 
Lastly, the effect of a $\hat{\sigma}^z_i$-measurement disentangle a single qubit on an entanglement cluster, collapsing it in the basis $\{|0\rangle,\ |1\rangle\}$.

\textbf{Simulation of measurement-only circuits.}
We summarize the algorithm presented in Ref.\cite{lang2020entanglement} to perform dynamics simulations.

$\bullet$ A vector $\mathbf{s}\in\mathbb{N}_0^L$ describes the state of the system, such that each site is described by a non-negative integer $s_i\in\mathbb{N}_0$; where $s_i=0$ indicates that site $i$ is unentangled from the rest of the system. Non-zero values $s_i=n>0$ show that site $i$ belongs to a cluster with any other site $j$ such that $s_j=n$, where there exists at least one more site $j$ satisfying this condition. The initial state is separable, therefore $\mathbf{s}=(0,\dots,0)$.

$\bullet$ The state is transformed depending on the operation. In a $\hat{\sigma}^z_i$-measurement on site $k$:the element $s_k$ is updated to $s_k'=0$, and the rest remain unchanged. An $\hat{\sigma}^x_i\hat{\sigma}^x_{i+1}$-measurement on $k$,$k+1$ transforms the vector as follows:

- Case 1: $s_k=0=s_{k+1}$, then the updated elements become $s'_k=s'_{k+1}=\text{next}(\mathbf{s})$. The function $\text{next}(\mathbf{s})$ returns the smallest integer unused in cluster labels of $\mathbf{s}$. 

- Case 2a: $s_k\neq 0$ and $s_{k+1}=0$; then $s_{k+1}'=s_k$.

- Case 2b: $s_k = 0$ and $s_{k+1} \neq 0$; then $s_k'=s_{k+1}$.

- Case 3: $s_k\neq 0$ and $s_{k+1} \neq 0$. With $s=\min(s_k,s_{k+1})$, set $s_l=s$ for all sites $l$ with $s_l=s_k$ or $s_l=s_{k+1}$.

The extraction of $\mathcal{F}_Q$ presented in Eq.~\eqref{eq_qfi_loop} is straightforward to calculate given $\mathbf{s}$ the final state of an individual disorder realization of the measurement-only circuits. By grouping integer values $v$ in $\mathbf{s}$ and saving their counts $c_v$, we can calculate
\begin{equation}
    \mathcal{F}_Q = c_0 + \sum_{v>0} c_v^2.
\end{equation}

\textbf{Bipartite entanglement analysis.} 
We provide a bipartite entanglement study of the phase-diagram in Fig.\ref{fig:phase_diagram}, by using the tripartite mutual information (TMI) of the final states
\begin{equation}
    \mathcal{I}_3 = S_A + S_B + S_C - S_{A \cup B} - S_{A \cup C} - S_{B \cup C} + S_{A \cup B \cup C},
    \label{eq:tmi}
\end{equation}
where $A,B,C$ indicate partitions of the system, and $S_K = -\Tr (\rho_K \log \rho_K)$  is the entanglement entropy or bipartite entanglement. Here, $\rho_K = \Tr_{\bar{K}}(\rho)$ is the partial trace over the complementary of the subsystem $K$, denoted by $\bar{K}$. We consider periodic boundary conditions and $A,B,C$ consecutive partitions of size $L/4$. This quantity is used in the literature~\cite{zabalo2020critical} as a convenient order parameter, with reduced finite-size drifts at the transition, to study its critical properties. \\
In the area-law phase with short-range correlations, the entanglement entropy is proportional to the boundary of the partition. Hence, all entropies with one boundary have $S_A=\mathcal A$, etc., while $S_{A\cup C}= 2\mathcal A$. Therefore, the contribution in Eq.\eqref{eq:tmi} compensates, and the tripartite mutual information vanishes $\mathcal{I}_3\to 0$.
Instead, cat states present long-range correlations and possess 
 $S_K=\log(2)$, independently of the partition, hence Eq.\eqref{eq:tmi} results in $\mathcal{I}_3=1$ \cite{klocke2024entanglementdynamicsmonitoredkitaev}.

\begin{figure}[t]
    \centering
    \includegraphics[width=0.4\textwidth]{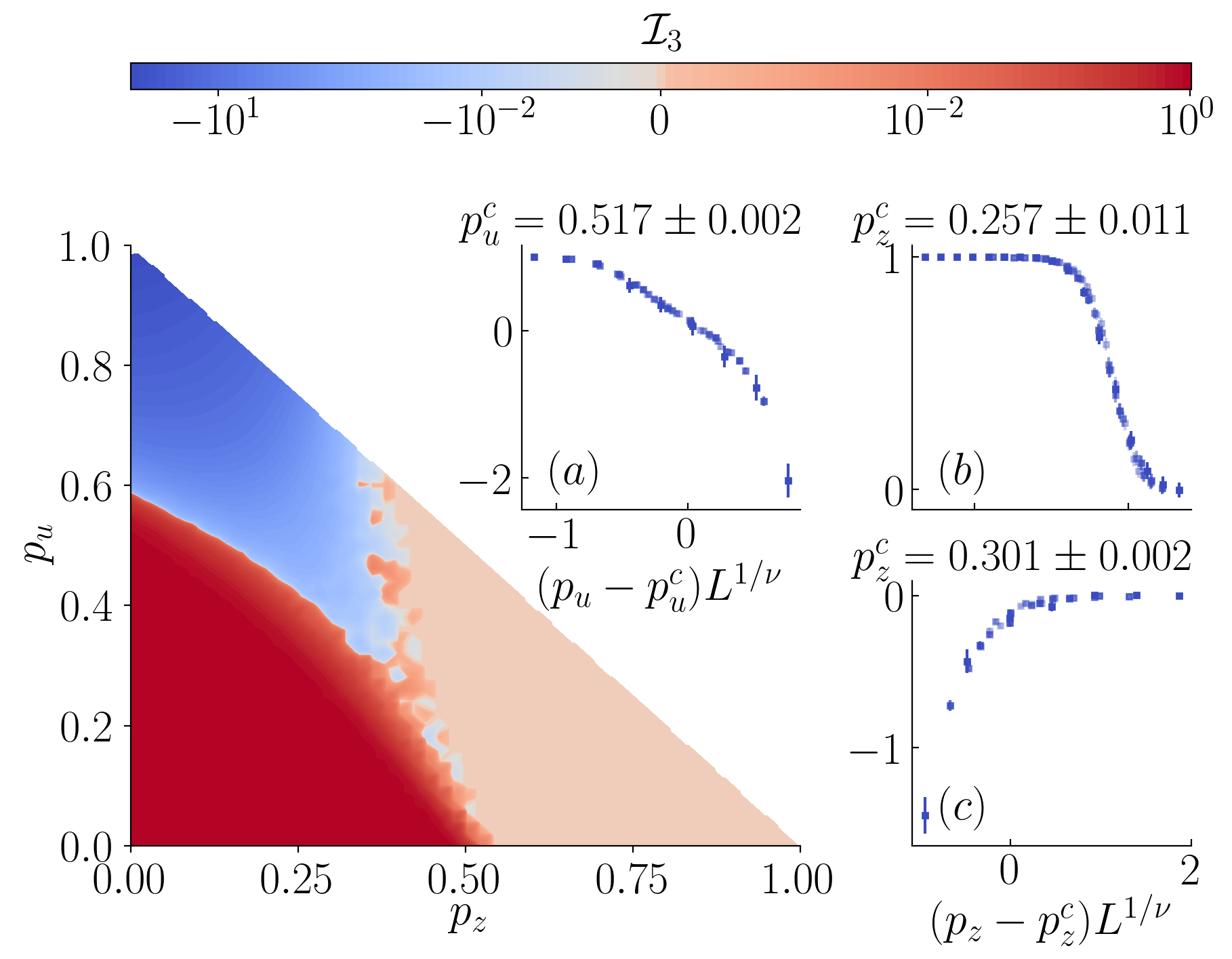}
    \caption{(main) Phase diagram of the TMI in the structured circuits parametrized by $(p_z,p_u)$. The data corresponds to a system size of $L=256$, and an average of 2000 disorder realizations. (a)-(c) Finite-size scaling of the lines studied in Fig.\ref{fig:phase_diagram} and respective critical probability, using system sizes $L=32,64,128,256,512$.}
    \label{fig:end_matt_i3}
\end{figure}

In Fig. \ref{fig:end_matt_i3} we show the TMI perspective of Fig. \ref{fig:phase_diagram}. We find three phases according to the different behavior of the TMI. In blue, when unitary operations are dominant, we have volume-law entanglement. Instead, when $\hat{\sigma}^z$ measurements predominate, we have an area-law phase with short-range correlations, shown in light red with $\mathcal{I}_3\sim 0$. In dark red we find a phase with long-range correlations, when $\hat{\sigma}_i^x\hat{\sigma}_{i+1}^x$ measurements predominate and $\mathcal{I}_3 = 1$.

Then, we focus on three lines $p_z=0.1$, $p_u=0.35$, and $p_{xx}=0.1$, and we accurately extract the critical properties of the entanglement phase transition through finite-size scaling on the data, as shown in Figs. \ref{fig:end_matt_i3}a-c:
\begin{equation}
    \mathcal{I}_3 (p, L) = F((p-p^c)L^{1/\nu})\ .
\end{equation}

\begin{figure}[H]
    \centering
    \includegraphics[width=0.4\textwidth]{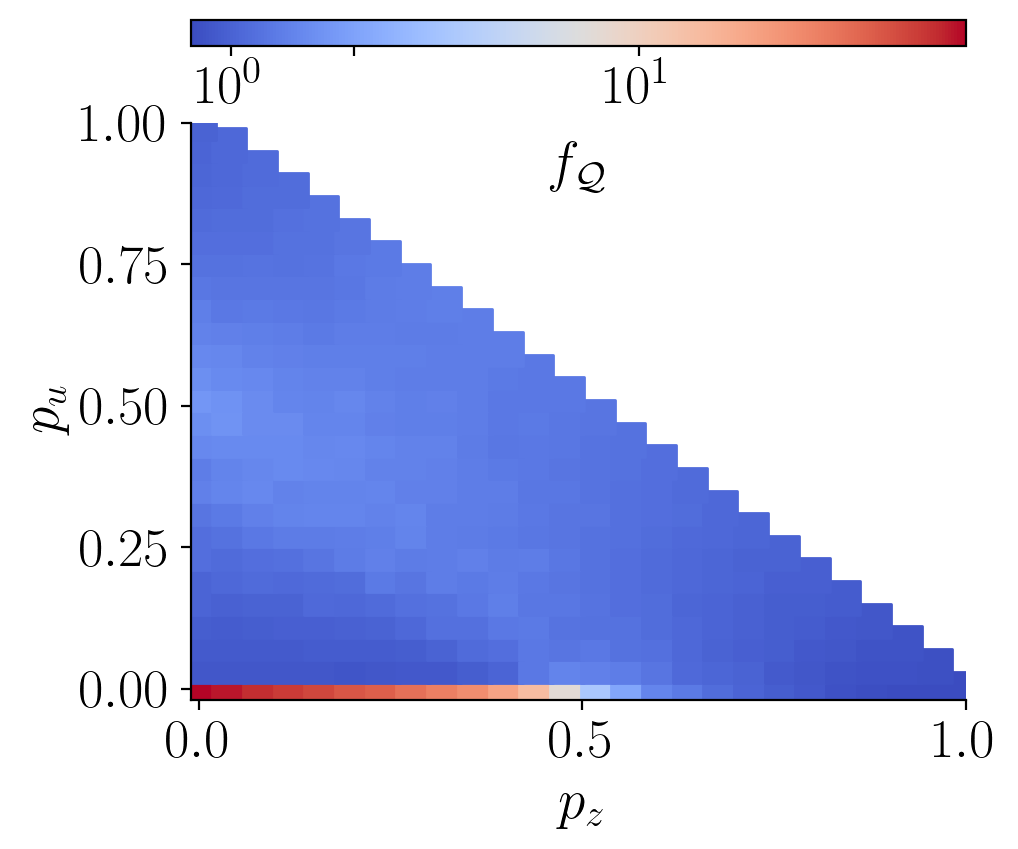}
    \caption{Phase diagram of the QFI in the circuit with random unitary operations (no symmetry), XX- and Z-measurements. The data corresponds to a system size of $L=64$, and an average of 100 disorder realizations.}
    \label{fig:end_matt_noSymm}
\end{figure}

\textbf{Structured circuits without symmetric unitary gates.} 
In Fig. \ref{fig:end_matt_noSymm}, we show the same phase diagram as in Fig.\ref{fig:phase_diagram}, but with the random unitaries $\hat U_{i, i+1}$ possessing no $\mathbb Z_2$ symmetry. In this case, the multipartite entanglement structure at $p_u>0$ becomes trivial. This result points out the relevance of the symmetry to observe different multipartite entanglement phases in random quantum circuits with $\hat{\sigma}_i^x\hat{\sigma}_{i+1}^x$ and $\hat{\sigma}^z$ measurements.

\end{document}